\documentclass[prd,tightenlines,nofootinbib,superscriptaddress]{revtex4}
\usepackage{amsfonts,amssymb,amsthm,bbm,amsmath}
\usepackage{color,psfrag}
\usepackage[dvips]{graphicx}

\newcommand{\C}{{\mathbb C}}
\newcommand{\N}{{\mathbb N}}
\newcommand{\R}{{\mathbb R}}

\newcommand{\cE}{{\mathcal E}}
\newcommand{\cF}{{\mathcal F}}
\newcommand{\cG}{{\mathcal G}}

\newcommand{\cL}{{\mathcal L}}
\newcommand{\cH}{{\mathcal H}}
\newcommand{\cM}{{\mathcal M}}

\newcommand{\cT}{{\mathcal T}}

\newcommand{\cP}{{\mathcal P}}

\newcommand{\cW}{{\mathcal W}}

\newcommand{\SU}{\mathrm{SU}}

\newcommand{\SO}{\mathrm{SO}}
\newcommand{\U}{\mathrm{U}}

\newcommand{\su}{{\mathfrak su}}

\renewcommand{\u}{{\mathfrak u}}

\newcommand{\be}{\begin{equation}}
\newcommand{\ee}{\end{equation}}
\newcommand{\beq}{\begin{eqnarray}}
\newcommand{\eeq}{\end{eqnarray}}
\newcommand{\bes}{\begin{eqnarray}}
\newcommand{\ees}{\end{eqnarray}}
\newcommand{\bea}{\begin{eqnarray}}
\newcommand{\eea}{\end{eqnarray}}
\newcommand{\nn}{\nonumber}

\newcommand{\mat} [2] {\left ( \begin{array}{#1}#2\end{array} \right ) }

\newcommand{\la}{\langle}
\newcommand{\ra}{\rangle}

\newcommand{\tr}{{\mathrm Tr}}
\newcommand{\f}{\frac}

\newcommand{\what}{\widehat}
\def\nn{\nonumber}
\def\pp{\partial}
\def\arr{\rightarrow}

\def\eps{\epsilon}
\def\tJ{\tilde{J}}
\def\tE{\tilde{E}}
\def\tF{\tilde{F}}
\def\tz{\tilde{z}}
\def\te{\tilde{e}}
\def\tg{\tilde{g}}

\def\bz{\bar{z}}

\def\vJ{\vec{J}}
\def\hJ{\hat{J}}
\def\hU{\hat{U}}
\def\hE{\hat{E}}
\def\hF{\hat{F}}
\def\vsigma{\vec{\sigma}}
\newcommand{\bra}[1]{\la #1 |}
\newcommand{\brar}[1]{[ #1 |}
\newcommand{\ket}[1]{ | #1\ra}
\newcommand{\ketr}[1]{ | #1]}
\newcommand{\braaket}[2]{\la #1 | #2 \ra}
\newcommand{\braaaket}[3]{\la #1 | #2|#3 \ra}

\newcommand{\ta}{\tilde{a}}
\def\tb{\tilde{b}}

\newcommand{\utwo}{\mathfrak{u}(2)}

\begin{document}

\title{Holonomy Operator and Quantization Ambiguities on Spinor Space}
\author{{\bf Etera R. Livine}, {\bf Johannes Tambornino}}
\address{Laboratoire de Physique, ENS Lyon, CNRS-UMR 5672, 46 All\'ee d'Italie, Lyon 69007, France}

\begin{abstract}

We construct the holonomy-flux operator algebra in the recently developed spinor formulation of loop gravity. We show that, when restricting to $\SU(2)$-gauge invariant operators, the familiar grasping and Wilson loop operators are written as composite operators built from the gauge-invariant `generalized ladder operators' recently introduced in the $\U(N)$ approach to intertwiners and spin networks. We comment on quantization ambiguities that appear in the definition of the holonomy operator and use these ambiguities as a toy model to test a class of quantization ambiguities which is present in the standard regularization and definition of the Hamiltonian constraint operator in loop quantum gravity.
\end{abstract}

\maketitle

\section{Introduction}

The recent development of spinor techniques for loop (quantum) gravity has led to various interesting applications, from a better understanding of the discrete geometries underlying spin network states to the derivation of Hamiltonian constraint operators encoding the dynamics prescribed by spinfoam models (see \cite{review_spinor} for a review).

To start with, the introduction of spinor variables allowed a compact reformulation of the loop gravity phase space \cite{twisted1,twisted2,spinor,spinor_johannes,spinor_conf}, with a clear geometrical interpretation as ``twisted geometries" generalizing the discrete Regge geometries. This became particularly relevant for the construction and interpretation of spinfoam models when analyzing the hierarchy of constraints to impose on arbitrary discrete space-time geometries in order to implement a proper quantum version of general relativity \cite{review_discrete}. Furthermore, following the generalization of these spinor variables to twistor networks allowing to describe a Lorentz connection \cite{lorentz1,lorentz2}, these spinor techniques (or actually upgraded to twistor techniques) allowed to explore and better understand the phase space structure underlying the discrete path integral defining the spinfoam amplitudes \cite{twistor_w,twistor_sf,spinor_dyn_w}.
Following a parallel but different line of research, this parametrization of the loop gravity phase space in terms of spinors naturally led to the definition of coherent states \cite{un2,un3,un4, un4_conf}, which allow a slight modification and a convenient re-writing of the spinfoam amplitudes. These techniques led to some exact results for the evaluation spinfoam amplitudes \cite{jeff, generating} or for the dynamics of spinfoam cosmology \cite{SFcosmo_merce}.
More generally, it is possible to describe and define the spinfoam amplitudes and 3nj symbols of the recoupling theory of spins ($\SU(2)$ representations) directly in terms of spinors and their quantization. At a semi-classical level, this allows to  derive and study their asymptotical behavior (at large spins) (e.g. \cite{asympt3nj}). At the full quantum level, this spinor techniques lead to differential equations and recursion relations satisfies by the spinfoam amplitudes, which are interpreted as the Hamiltonian constraints encoding the dynamics of the spin network states for the quantum geometry. This was indeed done explicitly for the BF theory spinfoam amplitudes \cite{recursion_spinor} (following the approach of \cite{dyn_3d} for 3d quantum gravity) and for the spinfoam cosmology amplitudes on the 2-vertex graph \cite{SFcosmo_merce}.
Finally, the use of spinor variables to parameterize the loop gravity phase space allowed a systematic study of the gauge invariant observables at the discrete level. This led to the identification of observables generating the basic $\SU(2)$-invariant deformations of intertwiners. We identified in particular $\u(N)$-subalgebra of observables (where $N$ is the valency of the intertwiner), which turned out powerful in the study of the intertwiner spaces \cite{un1} and the construction of appropriate coherent intertwiners \cite{un2,un3,un4} and in implementing symmetry reductions on fixed graph in order to define mini-superspace models for loop quantum gravity hopefully relevant for cosmology \cite{2vertex,SFcosmo_merce}.

\smallskip

The quantization of the spinor phase space is a priori rather straightforward since the complex variables are quantized as harmonic oscillators. The equivalence of this quantization scheme with the standard loop quantum gravity using spin network states was proved at the level of the Hilbert space on a fixed graph in \cite{spinor,spinor_johannes}. Nevertheless, all the relevant observables have not been consistently studied. Indeed, we have well studied the geometric observables (such as areas and angles) constructed from the triad classically and thus from the $\su(2)$ generators at the quantum level. However the holonomy operator has not yet been explicitly constructed in this context. The goal of the present short paper is to remedy this shortcoming.

We first remind the reader of the loop gravity phase space and its parametrization in terms of spinors. Then we describe the unambiguous quantization of the holonomy-flux algebra in terms of harmonic oscillators and holomorphic functions. This leads to a complete description of the grasping and holonomy operators of loop quantum gravity. In the spinorial picture these emerge both as composite operators built from some generalized ladder operators, $\hat{E}$'s and $\hat{F}$'s, which provide a complete set of $\SU(2)$-invariant operators living on each vertex and acting on interwiners \cite{un1,un2,2vertex,spinor}. This extends the work done for the classical phase space \cite{spinor} to the quantum realm.
Furthermore, we analyze some quantization- and operator-ordering ambiguities which are encountered in the definition of the holonomy operator on spinor space if we quantize it using the same technique as Thiemann's trick for the definition of the  Hamiltonian constraint operator in loop quantum  gravity \cite{thiemann_hamiltonian}. We show that it leads to an anomaly and we comment on the choice of quantization scheme.

\section{Loop Gravity with Spinors}
\label{review}

The Hilbert space of loop quantum gravity on a given oriented graph $\Gamma$ with $E$ edges\footnotemark is defined as the space of $L^2$ functions over $E$ copies of the $\SU(2)$ Lie group provided with the Haar measure, that is $\cH_\Gamma := L^2(\SU(2)^E, d^Eg)$.
\footnotetext{To be precise, the Hilbert space of loop quantum gravity arises as a sum over the individual graph-Hilbert spaces, or more precisely as an inductive limit,
$ \cH_{\rm LQG} := \overline{ \cup_\Gamma \cH_\Gamma / \sim   }$,
where $\sim$ denotes an equivalence relation between states living on different graphs and the completion in an appropriate topology is taken. See \cite{thiemannbook} for details.}.
This can be understood as a quantization of $(T^*\SU(2))^E$, namely one copy of the cotangent bundle $T^*\SU(2) \simeq \SU(2) \times \su(2)$ for each edge $e$ of the graph, which is usually parameterized with the couple $(g,J)$, where $g\in\SU(2)$ is the holonomy of the Ashtekar-Barbero connection along the edge $e$ and $J=\vJ\cdot\vsigma\in\su(2)\sim\R^3$ is related to the flux of the densitized triad through a surface dual to that edge. Furthermore, if $J$ is assumed to live on the target vertex of $e$ one can can use the group element $g$ to parallel-transport it to the source vertex of $e$ and obtain $J = -g \tJ g^{-1}$. The full Poisson algebra, attached to the edge $e$, is then given by
\beq
\label{T*SU(2)}
\{  J^i, J^j \} = \epsilon^{ijk} J^k,  && \{ \tilde{J}^i, \tilde{J}^k \}  = \epsilon^{ijk} \tilde{J}^k, \nn \\
\{ \vec{J}, g_{AB} \} = + \f i2 (\vec{\sigma}g)_{AB},
&& \{ \vec{\tilde{J}}, g_{AB} \} = - \f i2 ( g \vsigma)_{AB},  \nn \\
\{ g_{AB}  , g_{CD} \} = 0, && \{ J , \tilde{J} \} = 0  \, ,
\eeq
where $J$ and $\tJ$ are considered as 3-vectors and the $\SU(2)$ group element $g$ defined in the fundamental representation as a 2$\times$2 matrix with the indices $A,B=0, 1$.

\begin{figure}[h]
\begin{center}
\includegraphics[height=20mm]{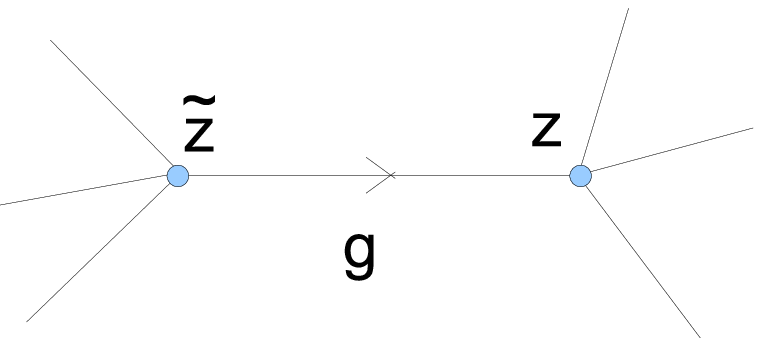}
\hspace*{15mm}
\includegraphics[height=30mm]{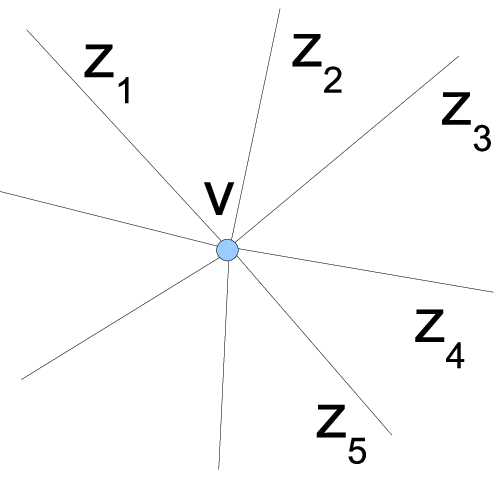}
\caption{ \label{spinor_network} On the left, an oriented edge of a spinor network carrying the holonomy $g$ which maps the spinor $\tz$ at its source vertex onto the spinor $z$ at its target vertex. On the right, a vertex $v$ of a spinor network with each of the attached edges carrying a spinor $z_i$}
\end{center}
\end{figure}
The next step is to impose $\SU(2)$ gauge invariance at the vertices $v$ of the graph. This is implemented by the (first class) closure constraints,
\be
\cG_v
\,\equiv\,
\sum_{e\ni v} J^e_v
\,=\,0\,,
\ee
which generates $\SU(2)$ transformation on both $J$'s and $g$'s. At the quantum level, it is taken into account by going from the Hilbert space $\cH_\Gamma = L^2(\SU(2)^E)$ to the gauge-invariant Hilbert space $\cH^o_\Gamma = L^2(\SU(2)^E/\SU(2)^V)$ where we have quotiented by the $\SU(2)$-action at all vertices.
The final step would be to implement the Hamiltonian constraints (corresponding to the invariance under space-time diffeomorphisms in the continuum theory). This issue is still-open and we will not discuss it here. We will focus on the kinematical structures of the theory.

\smallskip

Focusing on a single edge of the graph, it was shown in \cite{twisted1,twisted2,spinor,spinor_johannes,spinor_conf} that an alternative parameterization of $T^*\SU(2)$ is possible, in terms of two spinors  $\ket{z}, \ket{\tz} \in \C^2$. The spinors are interpreted as living at the source and target vertices of the edge, as illustrated on fig.\ref{spinor_network}.
A spinor $\ket{z}$ is an element of $\C^2$ and has components $z^A, \, A=0,1$. We denote its conjugate by $\bra{z} = (\bar{z}^0, \bar{z}^1)$ and its dual by $\ketr{z} = \epsilon \ket{\bar{z}}, \quad \epsilon = -i\sigma^2$. The space $\C^2$ has the standard positive definite inner product $\braaket{z}{w} := \bar{z}^0w^0 + \bar{z}^1 w^1$.


We further impose a matching constraint, $\cM := \braaket{z}{z} - \braaket{\tz}{\tz}$ enforcing the norms of the two spinors to be equal and generating a $\U(1)$ gauge invariance (under multiplication of the two spinors by opposite phases).
It can be shown that  $\C^2 \times \C^2$, the space of two spinors $\ket{z}$ and $\ket{\tz}$, reduces to $T^*\SU(2)$ by symplectic reduction with $\cM$ (see \cite{twisted1,twisted2,spinor_johannes} for details). The group and Lie-algebra variables are explicitly reconstructed as
\beq \label{fluxes}
 \vJ& = & \f12\braaaket{z}{\vec{\sigma}}{z}, \quad
 \tilde{J} = \f12\braaaket{\tz}{\vec{\sigma}}{\tz},  \nn\\
g & = & \frac{\ket{z}\brar{\tz} - \ketr{z}\bra{\tz}}{\sqrt{\braaket{z}{z}\braaket{\tz}{\tz}}} \, . \label{group_element}
\eeq
This relation obviously implies that $g\,|\tz]=|z\ra$ and thus reproduce the constraint $J = -g \tJ g^{-1}$.
Endowing $\C^2$ with the symplectic structure $\{ z^A, \bar{z}^B \} = -i \delta^{AB}$, one easily recovers the phase space \eqref{T*SU(2)}. In that sense, the spinors can be understood as (complex) Darboux-like coordinates for the space $T^*\SU(2)$ in which the symplectic structure \eqref{T*SU(2)} is trivial.

Here, let us stress a subtlety about the orientation of the edge. Indeed the group element introduced above maps one spinor onto the other, as $g\,|\tz]=|z\ra$ and $g\,|\tz\ra=-|z]$, so that its inverse is defined as $g^{-1}\,|z\ra=|\tz]$ and $g^{-1}\,|z]=-|\tz\ra$. This is slightly different from the change of orientation which would lead to a group element $\tg$ satisfying $\tg\,|z]=|\tz\ra$ and $\tg\,|z\ra=-|\tz]$ as we follow the definitions given above. The difference is actually just a sign flip:
$$
g^{-1}\,=\, \frac{|\tz]\la z| - |\tz\ra [z|}{\sqrt{\braaket{z}{z}\braaket{\tz}{\tz}}},
\qquad
\tg\,=\, \frac{|\tz\ra [z|-|\tz]\la z| }{\sqrt{\braaket{z}{z}\braaket{\tz}{\tz}}}
=-g^{-1}\,.
$$
This sign flip is not so important, but it is to be kept in mind. This sign ambiguity can be traced to the change from vector variables to spinor variables. Indeed, the Poisson algebra of the variables $(g,J)$ is unchanged under the change $g\leftrightarrow -g$, while the Poisson algebra of the variables $(g,z)$ will be affected (by a mere sign). This is simply because a 3-vector does not see the difference between the 3d rotations (in $\SO(3)$) induced by $g$ and $-g$, while these two $\SU(2)$ transformations act differently on a spinor.

\smallskip

This spinor parametrization provides a direct link between spin network states and discrete geometries and provides an interesting new perspective on loop quantum gravity \cite{spinor_johannes,spinor_conf,review_spinor},  spinfoam models \cite{un3,un4,un4_conf,twistor_sf,spinor_dyn_w}, quantum spinfoam cosmology \cite{2vertex,spinor,SFcosmo_merce},  topological BF-theory \cite{recursion_spinor} and group field theory \cite{gft_spinor}.

\smallskip

Now, turning the focus to a given vertex $v$ of the graph, one has one spinor variable $z_e$ per edge attached to $v$ (here we do not distinguish between the $z$'s and $\tz$'s), as illustrated on fig.\ref{spinor_network}. One can easily identify a complete set of $\SU(2)$-invariant observables, i.e commuting with the constraints $\cG_v$:
\be
E^v_{e_1e_2} := \la z_{e_1}| z_{e_2}\ra,
\qquad
F^v_{e_1e_2} := [ z_{e_1}| z_{e_2}\ra,
\qquad
\overline{F}^v_{e_1e_2} := \la z_{e_2}| z_{e_1}]\,.
\ee
These scalar products between spinors interestingly form a closed algebra \cite{un1,un2,spinor} and in particular the $E$-observable form a $\u(N)$ algebra (where $N$ is the number of edges attached to the vertex $v$) which is interpreted as generating the deformation of the intertwiner for fixed boundary area \cite{un1,un2,un0}. These observables then serve as basic building blocks for all gauge-invariant observables on a given graph $\Gamma$ and in particular allow to decompose at the classical level the holonomy observable into basic deformations of the spin(or) network \cite{spinor}. Here we will generalize this to the quantum level.

\section{The Holomorphic Representation}

The standard quantization scheme for  $T^*\SU(2)$ is to consider the Hilbert space $L^2(\SU(2))$, with the usual orthogonal basis given by the Wigner matrices $D^j_{mn}(g):=\la j,m|g|j,n\ra$ and with the holonomy $g$ acting by multiplication and the vector $X$ acting by derivation as the $\su(2)$ generators.
Here the choice of spinor variables to parameterize the phase space leads to a different polarization, but which has been shown to be unitarily equivalent to the standard quantization (e.g. \cite{spinor_johannes}).
The quantization of the canonical Poisson bracket
$\{ z^A, \bar{z}^B \}  = \{ \tz^A , \bar{\tz}^B \} = -i \delta^{AB}$
leads naturally to two copies of the Bargmann space $\cF_2 := L^2_{\rm hol}(\C^2, d\mu)$ of holomorphic, square integrable functions with a normalized Gaussian measure, where the spinors are naturally represented as the raising- and lowering operators of harmonic oscllators. Taking into account the (area) matching constraint $\hat{\cM} = 0$ one is led to the space
\be
\cH^{\rm spin} := \cF_2 \otimes \cF_2 / \U(1)\,.
\ee
A natural orthonormal basis of $\cH^{\rm spin}$ is given by holomorphic polynomials in two spinors,
\be    \label{P_basis}
\cP^j_{mn}(z,\tz) :=
\frac{(z^0)^{j+m}(z^1)^{j-m}}{\sqrt{(j+m)!(j-m)!}}
\frac{(-1)^{j+n}(\tz^1)^{j+n}(\tz^0)^{j-n}}{\sqrt{(j+n)!(j-n)!}} \, ,
\ee
where the spin $j \in \N/2$ gives the overall degree of the polynomial and $-j \leq \dots m,n \dots \leq +j$. These polynomials have a simple interpretation in terms of $\SU(2)$ representations:
\be
\cP^j_{mn}(z,\tz) = \la j,m|j,z\ra [j,\tz|j,n\ra
=\la j,m|j,z\ra \,\overline{\la j,n|j,\tz]}
\,,
\ee
where $|j,z\ra$ is the $\SU(2)$ coherent state labeled by the spinor $|z\ra$ while $|j,\tz]$ denotes the $\SU(2)$ coherent state labeled by the dual spinor $|\tz]=\eps\,|\bz\ra$ (see e.g. \cite{un2,un3,un4} for more details).
Orthonormality and completeness read as
\be
 \int d\mu(z) d\mu(\tz)\overline{\cP^{j}_{mn}(z,\tz)} \cP^{j'}_{m'n'}(z,\tz)   =   \delta_{jj'}\delta_{mm'}\delta_{nn'} \nn \; , \\
 \sum\limits_{ j \in \N/2} \sum\limits_{m,n=-j}^{j} \cP^{j}_{mn}(z_1, \tz_1) \overline{\cP^{j}_{mn}(z_2, \tz_2)} \nn \\  =  I_0(2\braaket{z_2}{z_1}\braaket{\tz_2}{\tz_1})\, ,\nn
\ee
where $d\mu$ is the Gaussian measure and $I_0(x)$ is the zeroth modified Bessel function of first kind which plays the role of the delta-distribution on $\cH^{\rm spin}$.

As explained in \cite{spinor_johannes,spinor_conf}, these holomorphic polynomials $\cP^j_{mn}$ are unitarily related to the standard Wigner matrices in $L^2(\SU(2))$, i.e. there exists a unitary map $\cT: L^2(\SU(2)) \rightarrow \cH^{\rm spin}$ mapping
\be \label{map_wigner}
D^j_{mn}(g) = \la j,m |g| j,n\ra
\quad\longmapsto\quad
(\cT D^j_{mn})(z, \tilde{z})
:=
\frac{1}{ \sqrt{2j+1}}\cP^{j}_{mn}(z, \tilde{z})
=
\frac{1}{ \sqrt{2j+1}}\la j,m|j,z\ra [j,\tz|j,n\ra
\,.
\ee
Considering the formula \eqref{group_element} for the holonomy $g$ in terms of the spinors $z,\tz$, it appears that this map $\cT$ extracts the holomorphic part of the group element $g$. The pre-factor in $\sqrt{2j+1}$ then ensures that this map still conserves the norms and scalar products, i.e. that it is unitary.
Details on this map and its properties can be found in \cite{spinor_johannes}.

\smallskip

The elementary operators on $\cH^{\rm spin}$ are the ladder-operators $[a, a^\dagger] = 1$ which directly correspond to the classical spinors $\{z  ,\bar{z}  \} = -i$. Holonomies $g$ and fluxes $J$ then emerge as composite operators.
Let us start with the basis (\ref{P_basis}) which decomposes as $\cP^j_{mn}(z,\tz) = e^j_{m}(z)\otimes \te^j_{n}(\tz)=e^j_{m}(z)\otimes \overline{e^j_{n}(|\tz])}$, where $e^j_m(z) := \frac{(z^0)^{j+m}(z^1)^{j-m}}{\sqrt{(j+m)!(j-m)!}}$ is the well known orthonormal Fock basis in $\cF_2$.
Each of the $z$ and $\tz$ spinors gets quantized into a set of two harmonic oscillator operators
\be \label{ho_commutators}
[\hat{a}^0,(\hat{a}^0)^\dagger] =  [\hat{a}^1,(\hat{a}^1)^\dagger] = 1,
\qquad
[\what{\ta^0},(\what{\ta^0})^\dagger] =  [\what{\ta^1},(\what{\ta^1})^\dagger] = 1\,.
\ee
These act as the usual raising- and lowering-operators on $\cF_2$ respectively:
\beq \label{osc_rep}
&&\bz^0\quad\longrightarrow\quad
(\hat{a}^0)^\dagger\,\, e^j_m(z) : = z^0 \,\,e^j_m(z)   = \sqrt{j+m+1}\,\, e^{j+\frac{1}{2}}_{m+\frac{1}{2} }(z) \nn \\
&&\bz^1\quad\longrightarrow\quad
(\hat{a}^1)^\dagger \,\,e^{j}_{m}(z) := z^1\,\, e^j_m(z)  = \sqrt{j-m+1}\,\, e^{j+\frac{1}{2}}_{m-\frac{1}{2} }(z) \nn \\
&&z^0\quad\longrightarrow\quad
\hat{a}^0 \,\,e^j_m(z) := \partial_{z^0}\,\, e^j_m(z)  = \sqrt{j+m}\,\, e^{j-\frac{1}{2}}_{m-\frac{1}{2} }(z) \nn \\
&&z^1\quad\longrightarrow\quad
\hat{a}^1 \,\,e^j_m(z) := \partial_{z^1}\,\, e^j_m(z)  = \sqrt{j-m}\,\, e^{j-\frac{1}{2}}_{m+\frac{1}{2} }(z) \; .
\eeq
At first, it might seem awkward that $\bz$ be quantized as the multiplication by $z$ while $z$ get becomes the differentiation $\pp_z$. To make it more normal, one should instead consider anti-holomorphic polynomials in the spinors $z$ and $\tz$, then $\bz$ would be the multiplication by $\bz$ and $z$ the differentiation $\pp_{\bz}$. This detail does not truly matter. What's important is the action of the operators on the basis states $e^j_m$ and how they shift the $j$ and $m$.

\smallskip

The quantization of the $\tz$-sector is carried out exactly as above for the $z$-sector. But since we act on slightly different wave-functions, the $\te^j_{n}(\tz)= \overline{e^j_{n}(|\tz])}$ instead of the $e^j_{m}(z)$, we will get different pre-factors and shifts in $j$ and $n$:
\beq \label{osc_rep_tl}
&&\overline{\tz}_0 \quad\longrightarrow\quad
(\what{\ta^0})^\dagger\,\, \te^{j}_{n}(\tz) := \tz^0 \,\,\te^j_n(\tz)
= \sqrt{j-n+1}\,\, \te^{j+\frac{1}{2}}_{n-\frac{1}{2} }(\tz) \nn \\
&&\overline{\tz}_1 \quad\longrightarrow\quad
(\what{\ta^1})^\dagger \,\,\te^j_n(\tz) : = \tz^2 \,\,\te^j_n(\tz)
= -\sqrt{j+n+1}\,\, \te^{j+\frac{1}{2}}_{n+\frac{1}{2} }(\tz) \nn \\
&&{\tz}_0 \quad\longrightarrow\quad
\what{\ta^0}\,\, \te^j_n(\tz) := \partial_{\tz^0}\,\, \te^j_n(\tz)
=  \sqrt{j-n}\,\, \te^{j-\frac{1}{2}}_{n+\frac{1}{2} }(\tz) \nn \\
&&{\tz}_1 \quad\longrightarrow\quad
\what{\ta^1} \,\,\te^j_n(\tz) := \partial_{\tz^1}\,\, \te^j_n(\tz)
= -\sqrt{j+n}\,\, \te^{j-\frac{1}{2}}_{n-\frac{1}{2} }(\tz) \; .
\eeq

\smallskip

The matching constraint is quantized as $\hat{\cM} = (\hat{a}^0)^\dagger \hat{a}^0 + (\hat{a}^1)^\dagger \hat{a}^1 - (\hat{\tilde{a}}^0)^\dagger \hat{\tilde{a}}^0 - (\hat{\tilde{a}}^1)^\dagger \hat{\tilde{a}}^1$, where we use the obvious notation $a,\tilde{a}$ (corresponding to $z, \tz$) to denote ladder-operators acting in the first and second copy of $\cF_2$ respectively. This generates a $\U(1)$-invariant on polynomials of $z$ and $\tz$ as expected and we check that $\hat{\cM}\cP^j_{mn} = 0$ for all basis elements of $\cH^{\rm spin}$. Then we require operators on the  space $\cH^{\rm spin}$, built from these ladder-operators on $\cF_2$,  to commute with the $\U(1)$-constraint $\hat{\cM}$.
Such invariant operators are the flux and holonomy operators, as we develop in the next section.

\section{Quantizing the Holonomy-Flux Algebra}
\label{formula}

Let us start with the flux-operators, corresponding to the quantization of the 3-vectors $J$:
\beq \label{flux_op}
\hat{J}^+ & = & (\hat{a}^0)^\dagger \hat{a}^1, \qquad \hat{J}^-  =  (\hat{a}^1)^\dagger \hat{a}^0 \nn \\
\hat{J}^3 & = & \frac{1}{2}[(\hat{a}^0)^\dagger \hat{a}^0 - (\hat{a}^1)^\dagger \hat{a}^1].
\eeq
The commutators are an exact representation of the classical Poisson brackets, i.e the $\hJ$ form an $\su(2)$-algebra:
\be
[\hat{J}^3, \hat{J}^{\pm}] = \pm  \hat{J}^{\pm}, \qquad  [\hat{J}^+, \hat{J}^- ] = 2 \hat{J}^3 \, .
\ee
This is the standard Schwinger representation for the $\su(2)$ Lie-algebra.
Their action on holomorphic polynomials is easily computed and reproduces the well-known action of the $\su(2)$ generators on basis states:
\beq \label{flux_action}
\hat{J}^+ \cP^{j}_{mn} & = & \sqrt{(j-m)(j+m+1)}\cP^j_{m+1,n}, \nn \\
\hat{J}^- \cP^j_{mn} & = & \sqrt{(j+m)(j-m+1)}\cP^j_{m-1,n}, \nn \\
\hat{J}^3 \cP^j_{mn} & = &   m \cP^j_{m n} \, .
\eeq
Another essential operator on $\cH^{\rm spin}$ is the area-operator $\hat{E}$ which arises as a quantization of the norm of the 3-vector $|\vJ| = \frac{1}{2}\braaket{z}{z}$,
\be
\hat{E} = \frac{1}{2}[(\hat{a}^0)^\dagger \hat{a}^0 + (\hat{a}^1)^\dagger \hat{a}^1]
\ee
which is diagonalized in the standard basis (\ref{P_basis}) such that
\be \label{action_E}
\hat{E}\cP^j_{mn} = j \cP^j_{mn}
\ee
and commutes with $\hat{J}^{\pm}, \hat{J}^3$. The geometric interpretation of this operator $\hE$ is that it gives the area carried by that edge.
The set $\{ \hat{J}^\pm, \hat{J}^3, \hat{E} \}$ forms a $\utwo$-algebra.

\smallskip

The $\tz$-sector differs slightly from the formulas above. The quantization of the 3-vectors is carried out exactly the same way and the expression of the operators $\tJ$ in terms of the raising and lowering operators remains the same. Nevertheless the action on basis states gets a sign flip:
\beq \label{flux_action_tl}
\what{\tJ^+}\, \cP^{j}_{mn} & = & -\sqrt{(j+n)(j-n+1)}\,\cP^j_{m,n-1}, \nn \\
\what{\tJ^-}\, \cP^j_{mn} & = & -\sqrt{(j-n)(j+n+1)}\,\cP^j_{m,n+1}, \nn \\
\what{\tJ^3}\, \cP^j_{mn} & = &   -n \,\cP^j_{m n}, \nn \\
\what{\tE}\,\cP^j_{mn} &= & j \cP^j_{mn} \,=\,\hat{E}\,\cP^j_{mn}\, .
\eeq
Nevertheless the operators $\what{\tJ}$ still form the expected $\su(2)$ algebra without any sign flip (this is actually the complex conjugate representation of $\su(2)$ compared to the $z$-sector).

\smallskip

To derive the action of the holonomy operator, we start by its action on the Wigner matrices in $L^2(\SU(2))$ and pull back to $\cH^{\rm spin}$ using the unitary map $\cT$ as in (\ref{map_wigner}).
Indeed, on $L^2(\SU(2))$ we know that the holonomy operators $\hat{g}_{AB}$ (here taken as the matrix elements of the group element $g$ in the fundamental representation of $\SU(2)$) acts simply by multiplication, that is
\be
\hat{g}_{AB}\psi(g) \, = \, g_{AB}\psi(g)\quad \forall \psi \in L^2(\SU(2)) \, .
\ee
Using the $\SU(2)$-recoupling theory, the action of the holonomy is easily computed. It is convenient to switch from the indices $A,B=0,1$ to indices $\alpha,\beta=\pm\f12$. We get after the pull-back:
\be \label{g_holomorphic_1}
\hat{g}_{\alpha\beta} \cP^j_{mn}
\,=\,
\frac{4\alpha\beta}{2j+1}\sqrt{(j+2\alpha m+1)(j+2\beta n+1)} \cP^{j+\frac{1}{2}}_{m+\alpha, n+\beta}
\,+\,
\frac{1}{2j+1}\sqrt{(j-2\alpha m)(j-2\beta n)} \cP^{j-\frac{1}{2}}_{m+\alpha, n+\beta} \, .
\ee
Note that the precise pre-factor $\frac{1}{2j+1}$ is crucial to ensure that the classical Poisson algebra relations (\ref{T*SU(2)}) are correctly implemented on $L^2(\SU(2))$.
In particular, one can apply this formula to the special case when we act with the character $\chi_{\f12}$ on the character $\chi_j$:
\be
\label{hol_action}
\what{\chi_{\f12}(g)}\,\, \chi_j
\,=\,
\sum_{alpha=\beta}\hat{g}_{\alpha\beta}\,\,\sum_{m=n}P^j_{mn}
\,=\,
\sum_m P^{j+\f12}_{mm} + P^{j-\f12}_{mm}
\,=\,
\chi_{j+\f12}+\chi_{j-\f12}\,,
\ee
as expected.
Instead of using the formulas from $\SU(2)$ recoupling, we can follow our quantization rules from the classical expression \eqref{group_element} of the holonomy:
\beq
&g=
\f{1}{\sqrt{\braaket{z}{z}\braaket{\tz}{\tz}}}
\mat{cc}{\bz^1\overline{\tz^0}-z^0\tz^1 & z^0\tz^0+\bz^1\overline{\tz^1} \\ -\bz^0\overline{\tz^0}-z^1\tz^1 & z^1\tz^0-\bz^0\overline{\tz^1}}& \nn\\
&\quad\longrightarrow\quad
\hat{g}
=\mat{cc}{g_{--} &g_{-+} \\ g_{+-}&g_{++}}
=\mat{cc}{(\hat{a}^1)^\dagger (\what{\ta^0})^\dagger - \hat{a}^0 \what{\ta^1}
& \hat{a}^0 \hat{\ta}^0 + (\hat{a}^1)^\dagger (\what{\ta^1})^\dagger \\
-\hat{a}^1 \hat{\ta}^1 - (\hat{a}^0)^\dagger (\what{\ta^0})^\dagger
& \hat{a}^1 \what{\ta^0}-(\hat{a}^0)^\dagger (\what{\ta^1})^\dagger}
\frac{1}{2 \hat{E} + 1} \,.&
\label{g_op_1}
\eeq
And one finds the exact same action as above when computed using the recoupling of $\SU(2)$ representations.

The polynomial part is straightforwardly quantized and the only subtlety is the pre-factor $\frac{1}{\sqrt{\braaket{z}{z}\braaket{\tz}{\tz}}}$ that gets regularized and quantized as $(2\hat{E} + 1)^{-1}$. This regularization, which makes the non-polynomial part a well-defined operator on all of $\cH^{\rm spin}$ (the a priori expression $(2\hat{E})^{-1}$ diverges on the state $\cP^0_{mn}$ for $j=0$), seems a bit ad hoc at the first glance. However, there are two reasons which justify this choice: first, this is the only regularization which ensures that $\hat{g}$ acts on $L^2(\SU(2))$ as required (after acting with the map $\cT$). Second, even without knowing about the unitary mapping between $L^2(\SU(2))$ and $\cH^{\rm spin}$, one would have constructed \emph{the same} operator $\hat{g}$ by  starting with an arbitrary regularization $\frac{1}{f(\hat{E})}$ of the non-polynomial part of the group element such that it is well-defined on all states in $\cH^{\rm spin}$ and then demand that the commutator $[\hat{g}_{AB}, \hat{g}_{CD}] = 0$ is implemented with no-anomaly. This condition selects the regularization chosen here.

\smallskip

Now that we have constructed the operators  $\hat{J},\,  \hat{\tilde{J}}$ and  $\hat{g}_{AB}$, we still have to check the final commutation relations between them in order to conclude that the Poisson algebra (\ref{T*SU(2)}) is represented correctly on $\cH^{\rm spin}$. The Poisson brackets between $J$'s and the $g$'s can be re-written explicitly as:
\be
\{ J^3\,, \,g_{\alpha, \beta} \} \,= \,-i\,\alpha \,g_{\alpha ,\beta},
\qquad
\{ J^\pm\,, \,g_{\alpha ,\beta} \} \,= \,i \,(\frac{1}{2} \mp \alpha)\, g_{-\alpha ,\beta}\,,
\ee
which gets quantized as:
\be
[\hat{J}^3\,, \,\hat{g}_{\alpha, \beta}] \,= \,\alpha \,\hat{g}_{\alpha ,\beta},
\qquad
[ \hat{J}^\pm\,, \,\hat{g}_{\alpha ,\beta} \}\, =\, -(\frac{1}{2} \mp \alpha)\, \hat{g}_{-\alpha ,\beta}\,.
\ee
It is straightforward that these commutators are satisfied by the operators as we have defined above.

\section{Quantization Ambiguity and Thiemann's Trick}

As stated above, the ordering ambiguities in the holonomy-operator on $\cH^{\rm spin}$ are strongly restricted by demanding that the classical Poisson algebra (\ref{T*SU(2)}) be represented non-anomalously. A conceptually similar, but mathematically much more involved problem occurs in the definition of the Hamiltonian constraint, which generates the quantum dynamics in loop quantum gravity.
%
The issue is that the classical Hamiltonian constraint is polynomial but for a pre-factor given by he inverse square-root of the determinant of the triad, which does not have a clear unambiguous quantization.
Thiemann \cite{thiemann_hamiltonian} was the first to propose a mathematically well-defined quantum operator corresponding to the classical Hamiltonian constraint. His construction relies strongly on some classical Poisson-identities, reexpressing the inverse square root of $\det \cE$ which appears in the classical Hamiltonian constraints as
\be \label{poisson_identity}
\frac{\epsilon_{abc}\epsilon_{jkl}\cE^b_k \cE^c_l}{4 \sqrt{|\det \cE|}} = \{  A_a^j, V \}
\ee
where $V = \int\limits_{\Sigma}d\sigma \sqrt{|\det \cE|}$ is the total volume of the spatial manifold $\Sigma$. The basic variables are  the Ashtekar-connection  $A_a^j$ and the densitized triad $\cE^a_j$, which is  related to the (inverse) spatial metric as $q^{ab} = \cE^a_j \cE^b_j/ |\det \cE|$. This Poisson-identity is used to reformulate the (Euclidean part of the) classical Hamiltonian constraint as
\be
H_{\rm Eucl} = \frac{\tr(F_{ab} \cE^a \cE^b)}{\sqrt{|\det \cE|}} = \tr (  F \wedge \{ A , V \} ) .
\ee
and the corresponding quantum operator is then defined as\footnote{There are some more steps to be taken, a regularization is chosen adapted to the graph, connection $A$ and curvature $F$, which do not exist as operators on the Hilbert space of loop gravity, are approximated by holonomies, etc. However, a key ingredient in the definition of the constraint, which amounts to choosing a particular operator ordering, is the Poisson identity (\ref{poisson_identity}). The main argument in favor of this operator ordering is that the classical volume functional $V$ has a well defined quantum analogue in loop gravity.} $\hat{H}_{\rm Eucl} := \frac{1}{i\hbar}  \tr(\hat{F}\wedge [\hat{A}, \hat{V}])$.
This definition, using the Poisson-identity  (\ref{poisson_identity}) as a way to regulate a possibly diverging non-polynomial expression is rather non-standard and its physical and mathematical consistency has not been checked intensively so far. Whether the  classical hypersurface deformation algebra, encoding general relativity's invariance under diffeomorphisms, is represented non-anomalously remains an open issue \cite{lewandowski_marolf_97, glmp_97}.
To analyze the fate of the hypersurface deformation algebra in the full theory is rather difficult conceptually and technically. However, the spinorial formalism described in this article can be used to model a quantization based on such Poisson-identities in a much simpler setting. Indeed the holonomy $g$ contains a similar pre-factor, given by the inverse square-root of the product of the norms of the spinors. It is possible to use a similar trick to re-absorb this pre-factor and generate it through a Poisson bracket. But in our (much) simpler framework, we know the exact quantization of the holonomy $\hat{g}$, so we can test if such Poisson-identities lead more or less to the correct quantization or not.

More precisely, we consider the Poisson-bracket of a spinor $\ket{z}$ with the square-root of the total area $E = \frac{1}{2}\braaket{z}{z}$, which allows to generate an inverse square-root of the norm of $z$:
\be \label{poisson_identity_spinors}
\{|z\ra,\sqrt{E}\}
\,=\,
\f1{2\sqrt{E}}\,\{|z\ra,E\}
\,=\,
\f{-i\sqrt{2}}{4}\,\f{|z\ra}{\sqrt{\la z|z\ra}},
\qquad
\{|z],\sqrt{E}\}
\,=\,
\f{+i\sqrt{2}}{4}\,\f{|z]}{\sqrt{\la z|z\ra}},
\ee
Using the same Poisson-identities for the spinor $\tz$, we can therefore write the classical group element (\ref{group_element}) as
\be
g\, =\, -8\, \big{\{} \sqrt{\tilde{E}}\,,\, \big{\{}  \sqrt{E}\, ,\, \ket{z}\brar{\tz} - \ketr{z}\bra{\tz} \,\big{\}}\big{\}} \, .
\ee
Similarly to the definition of the Hamilton constraint operator in loop quantum gravity we can now promote this identity to the definition of a holonomy operator  by replacing the Poisson brackets by commutators, $\{\cdot,\cdot\} \rightarrow \,-i [\cdot ,\cdot]$. We simply substitute the classical phase space functions $\sqrt{E}$, $\sqrt{\tilde{E}}$ and  $\ket{z}\brar{\tz} - \ketr{z}\bra{\tz}$ with the corresponding well-defined\footnote{$\hat{E}$ is a well defined, positive operator  (\ref{action_E}). Its spectrum is bounded from below by $0$, therefore the operator $\widehat{\sqrt{E}}$ is well-defined on $\cH^{\rm spin}$ and acts on the basis $\cP^j_{mn}$ as $\widehat{\sqrt{E}}\,\cP^j_{mn}\, =\, \sqrt{j}\,\,\cP^{j}_{mn}$. Furthermore, because all functions in $\cH^{\rm spin}$ are $\U(1)$-invariant so that $\hat{\tilde{E}} = \hat{E}$ as operators on $\cH^{\rm spin}$.} operators on $\cH^{\rm spin}$. As a result we obtain the new definition:
\be
\what{ g'} \,\equiv \,8\, \big{[} \widehat{\sqrt{E}}\,,\, \big{[} \widehat{\sqrt{E}} \,,\, \ket{\hat{a}^\dagger}\brar{\hat{\tilde{a}}} - \ketr{\hat{a}^\dagger}\bra{\hat{\tilde{a}}} \,\big{]}\big{]} \, .
\ee
It is straightforward to compute its action on basis states:
\beq
{\what{g'}_{\alpha\beta}} \cP^j_{mn}
&=&
8 (\sqrt{j+\frac{1}{2}} - \sqrt{j})^2 \sqrt{(j+2\alpha m+1)(j+2\beta n+1)}\cP^{j+\frac{1}{2}}_{m+\alpha,n+\beta} \nn\\
&&+ 4\alpha\beta \,8(\sqrt{j} - \sqrt{j-\frac{1}{2}})^2 \sqrt{(j-2\alpha m)(j+2\beta n)} \cP^{j-\frac{1}{2}}_{m+\alpha, n+\beta} \, .
\eeq
Note that the action of this operator is very similar to (\ref{g_holomorphic_1}), but for the different  pre-factors in $j$ in front of each term. In the asymptotic limit of large spins $j$, we do recover that the pre-factors above give back $1/(2j+1)$ as expected at leading order. But for small $j$'s, the pre-factors differ, which means that the quantization of the holonomy is clearly different for small spins, i.e. close to the Planck scale.
Moreover, as stated above, the exact form of the combinatorial pre-factors in (\ref{g_holomorphic_1}) \emph{is} important in order to obtain a non-anomalous representation of the classical Poisson-algebra (\ref{T*SU(2)}). With the latter choice derived from quantizing the Poisson-identities,
\be
[ \what{g'}_{\alpha\beta}, \what{g'}_{\gamma\delta} ] \cP^j_{mn} \neq 0  \, ,
\ee
which contradicts the fact that any two holonomy operators should Poisson-commute.

\smallskip

In general, if we define the holonomy operator acting as:
\be
\what{G}_{\alpha\beta}\, \cP^j_{mn}
\,\equiv\,
{4\alpha\beta}\,f_+(j)\,\sqrt{(j+2\alpha m+1)(j+2\beta n+1)} \cP^{j+\frac{1}{2}}_{m+\alpha, n+\beta}
\,+\,
f_-(j)\,\sqrt{(j-2\alpha m)(j-2\beta n)} \cP^{j-\frac{1}{2}}_{m+\alpha, n+\beta} \,,
\ee
where $f_+(j)$ and $f_-(j)$ are the $j$-dependent pre-factors corresponding to the quantization of the inverse square-root of the norms of the spinors, we can compute the resulting commutator. We only give the commutator between $\what{G}_{++}$ and $\what{G}_{--}$ for the sake of simplicity (to avoid a mess with the indices), but all the commutators can be computed similarly:
\be
[\what{G}_{++},\what{G}_{--}]\,\cP^j_{mn}
\,=\,
2(m+n)\,\bigg{[}
j\,f_-(j)f_+(j-\f12)\,-\,(j+1)\,f_+(j)f_-(j+\f12)
\bigg{]}
\,\cP^j_{mn}\,.
\ee
It is fairly easy to check that this factor vanishes for our quantization $\hat{g}$, when $f_-=f_+=(2j+1)^{-1}$. On the other hand, for the quantization using the Thiemann-like trick, we get for $[\what{g'}_{++},\what{g'}_{--}]$:
\be
\bigg{[}
j\,f_-(j)f_+(j-\f12)\,-\,(j+1)\,f_+(j)f_-(j+\f12)
\bigg{]}
\quad\underset{j\arr\infty}{\sim}
\f{1}{8}\,\f1{j^3}\,.
\ee

\smallskip

Therefore we conclude that a quantization based on the Poisson-identity (\ref{poisson_identity_spinors}) does not give the desired result and leads to an anomaly in the algebra at the quantum level. Defining an operator via Poisson-identities of the kind (\ref{poisson_identity_spinors}) amounts to a specific choice of operator ordering in the quantum theory. In the simple test case considered here, we have shown that this particular operator ordering does not lead to a proper quantum representation of the classical Poisson algebra.

This is a standard with the quantization of non-polynomial observables. Having a closed algebra of observables at the classical level guides us here to choose the correct quantization and operator ordering.

The quantum dynamics of loop quantum gravity (and loop quantum cosmology, which is often used as a finite dimensional toy model) relies substantially on the Poisson-identity (\ref{poisson_identity}) which is, at least in spirit, very similar to the one tested here. Extrapolating from our results to the case of loop quantum gravity, it would not be surprising if a similar inconsistency would show up in the quantization of the Hamiltonian constraint. To clarify this issue we think it could be helpful to study further toy models based on Poisson identities of the type (\ref{poisson_identity}) and check them for internal consistency.

\section{Graspings and Volume Operator}

So far we have restricted our attention to operators that were defined on a single edge of a spin network, namely the holonomy- and flux-operators (\ref{flux_op}) and (\ref{g_op_1}) acting on $\cH^{\rm spin}$.
Now we would like to discuss operators acting on the full spin network, and thus taking into account the $\SU(2)$ gauge invariance at the vertices of the graph. The Hilbert space on an arbitrary graph $\Gamma$ is $\cH_{\Gamma}=L^2(\SU(2)^{\times E}/\SU(2)^{\times V})$ in terms of the total number of edges $E$ and the total number of vertices $V$. It can be recast in the spinor framework as \cite{spinor,spinor_johannes}:
\beq
\cH_{\Gamma}
&=&
L^2(\SU(2)^{\times E}/\SU(2)^{\times V})
\,=\,
\left( \mathop{\otimes}\limits_{e=1}^{E} \cH^{\rm spin}_e \right) / \SU(2)^V \nn\\
&=&
\mathop{\otimes}\limits_{e=1}^{E} (\cF_2 \otimes \cF_2) / \U(1)^E / \SU(2)^V
\,=\,
\left[\mathop{\otimes}\limits_{v=1}^V\left(\mathop{\otimes}\limits_{i=1}^{N_v} \cF_2\right) / \SU(2) \right] / \U(1)^E
\eeq
where $i = 1,..,N_v$ labels the edges attached to a given vertex $v$.
The initial definition focuses on degrees of freedom attached to the edges of the graph -the group element $g_e$- up to the $\SU(2)$ gauge invariance at the vertices. The spinor framework allows to break the degrees of freedom on each edge into two pieces -the two spinors $z_e$ and $\tz_e$- attached to its source and target vertices, up to a $\U(1)$ gauge invariance along the edge allowing to glue the two spinors (by imposing that their norm be equal). In the end, this allows to factorize the Hilbert spaces around each vertex and to re-write the full Hilbert space as encoding degrees of freedom attached to each vertex up to the $\U(1)$ gauge invariance on all edges. This is natural from the point of view that spin networks are made of intertwiners living at each vertex of the graph, or equivalently at the classical level that spinor networks can be interpreted geometrically as polyhedra dual to each vertex and glued together along the edges \cite{un1,twisted1,twisted2,spinor}. This is at the heart of the $\U(N)$ framework for spin networks \cite{un1,un2,spinor,un3,un4}, where the space of intertwiners at each vertex $v$ is identified as living in an irreducible representation of the unitary group $\U(N_v)$ (which depends on the total area around the vertex $v$).

\smallskip

Indeed, now focusing on a vertex $v$ of the graph, we have $N$ edges attached to it (we dropped the index $v$ off $N_v$), and thus $N$ spinors $z_i$. As we have seen earlier in section \ref{review}, we have a set of $\SU(2)$-invariant observables at the classical level given by the scalar between those spinors and their dual:
$$
E^v_{ij} := \la z_{i}| z_{j}\ra,
\qquad
F^v_{ij} := [ z_{i}| z_{j}\ra,
\qquad
\overline{F}^v_{ij} := \la z_{j}| z_{i}]=-\la z_{i}| z_{j}]\,,
$$
where the $E$-matrix is Hermitian, $E_{ij}=\overline{E_{ji}}$ and the $E$-matrix anti-symmetric, $F_{ij}=-{F_{ji}}$
At the quantum level, we have working on the Hilbert space
$\cH_v :=\left(\mathop{\otimes}\limits_{i=1}^{N} \cF_2\right) / \SU(2)$. The quantization of those observables is straightforward:
\be \label{E_F_def}
\hat{E}_{ij} := (\hat{a}^0_i)^\dagger \hat{a}^0_j +  (\hat{a}^1_i)^\dagger \hat{a}^1_j,
\qquad
\hat{F}_{ij} := \hat{a}^0_i\hat{a}^1_j - \hat{a}^1_i\hat{a}^0_j
\qquad
\hat{F}_{ij}{}^\dagger := \hat{a}^0_i{}^\dagger\hat{a}^1_j{}^\dagger - \hat{a}^1_i{}^\dagger\hat{a}^0_j{}^\dagger
\,.
\ee
These operators are now the basic building blocks for all $\SU(2)$-invariant operators acting on spin networks. The $\U(N)$ formalism is based on the fact that the $\hE_{ij}$ operators form a closed $\u(N)$ algebra \cite{un0,un1}.

\smallskip

Now we are interested in the $\SU(2)$-invariant version of the flux and holonomy operators, studied in the previous section. We construct in the present section the grasping operators around a given vertex $v$, as polynomials in the operators $E^v$  and $F^v$. In the next section, we will deal with the holonomy operators, defined around closed loops of the graph. They will involve polynomials of the operators $E^v$  and $F^v$ (for all vertices $v$ around the loop), up to the same norm factors that appeared for the quantization of the group element on a single edge. We will pay special attention to those factors and associated operator ordering.

\smallskip

Coming back to a single vertex, the operators $\hE^v_{ij}$ and $\hF^v_{ij}$ acts on the couple of edges $i$ and $j$ attached to the vertex $v$.
The $\hat{E}$-operators shift a quantum of area from one edge to another, while the $\hat{F}$- and $(\hat{F}^\dagger)$-operators respectively annihilate and create a quantum of area on both edges. In this sense these operators can be regarded as generalized creation and annihilation operators.
They are invariant under the action of $\SU(2)$ at the vertex. They are not however invariant under the $\U(1)$-symmetry on the edges and thus do not qualify as operators on the fully gauge invariant Hilbert space $\cH_\gamma$.

On the other hand, the natural observables invariant under both $\U(1)$ and $\SU(2)$ symmetries are the scalar combination of the 3-vectors, i.e the scalar product $\vec{J}_i \cdot \vec{J}_j$ and higher order combinations involving vector products such as $(\vec{J}_i \wedge \vec{J}_j)\cdot \vec{J_k}$. All these observables can actually be written in terms of the $E$ and $F$ observables as polynomials (of the same in $E,F$ as in the $J$'s).
For instance, starting with the scalar product observable on a single edge, this gives the squared area carried by that edge, which is easily translated in the $E$-observables:
\be
|\vec{J}_i|^2
\,=\,
\vec{J}_i \cdot \vec{J}_i
\,=\,
E_{ii}^2\,.
\ee
This relation also holds at the quantum level, except for a correction term accounting for the quantum ordering \cite{un0,un1}:
\be
\vec{\hJ}_i\cdot\vec{\hJ}_i
\,=\,
\hE_{ii}\,(\hE_{ii}+1)\,.
\ee
Such correction terms lead to ordering ambiguities in the area spectrum of loop quantum gravity. For instance, one can define the area directly as the operator $\hE_{ii}$, with spectrum $j$, or as the squareroot of the $\SU(2)$ Casimir operator $\sqrt{\vec{\hJ}^2_i}$, with spectrum $\sqrt{j(j+1)}$. Such ambiguities appear crucial in solving some of the (second class) constraints  in loop gravity or spinfoam, for instance in the construction of the EPRL spinfoam model \cite{eprl}.
Notice nevertheless that the operators $\hE_{ii}$ and  $\sqrt{\vec{\hJ}^2_i}$ are different and do not have the same commutation relation with the other observables. Therefore selecting the particular Poisson bracket that we want to keep (without anomaly) as commutators at the quantum level would select one particular ordering over all others.
The same is easily done for the scalar product observables between two different edges \cite{un1,2vertex}:
\beq
&&\vec{J}_i \cdot \vec{J}_j
\,=\,
- \frac{1}{2}|F_{ij}|^2 + \frac{1}{4}{E}_{ii}{E}_{jj}
\,=\,
\frac{1}{2}|{E}_{ij}|^2 - \frac{1}{4}{E}_{ii}{E}_{jj}\nn\\
&\longrightarrow\quad&
\vec{\hJ}_i \cdot \vec{\hJ}_j
\,=\,
- \frac{1}{2}\hF^\dagger_{ij}\hF_{ij} + \frac{1}{4}{E}_{ii}{E}_{jj}
\,=\,
\frac{1}{2}\hat{E}_{ij}\hat{E}_{ji} - \frac{1}{4}\hat{E}_{ii}\hat{E}_{jj} - \frac{1}{2}\hat{E}_{ii}\,.
\eeq
We also look at the cubic operator $\hU_{ijk}\equiv -i[\vec{\hJ}_i\cdot\vec{\hJ}_j,\vec{\hJ}_i\cdot\vec{\hJ}_k]=\eps_{abc}\hJ^a_i\hJ^b_j\hJ^c_k$. For a 3-valent vertex, this operator vanishes due to the $\SU(2)$ gauge-invariance. It is non-trivial for a 4-valent vertex and actually defines the squared volume  operator (up to a numerical factor) (see e.g. \cite{barbieri} for the geometrical interpretation or \cite{thiemann_vol,thiemann_vol2} for a study of this operator in loop quantum gravity and more recently \cite{marzuoli2013}). The peculiarity of this operator is that it is not positive (its spectrum is real but symmetric under change of sign), so it is non-trivial to define its square-root. One can very naively take its absolute value in a basis which diagonalize it, but this seems an ad hoc definition weaken by the fact that we do not know explicitly the exact spectrum and eigenstates of the operator $U$. Nevertheless, it is the only well-defined proposal for a volume operator. It would seem better suited to identify the positive and negative modes of the operator but this turns out more complicated and it is not yet achieved\footnotemark. For vertices with valency larger or equal to 5, the squared volume operator is defined by adding the operators $U_{ijk}$ over all (oriented) triplets of edges attached to the vertex. This operator turns out to be easily written in terms of the operators $E$ or $F$. After a little algebra, we obtain:
\be
\hU_{ijk}
\,=\,
\f{-i}{4}\,\left(
\hE_{ij}\hE_{jk}\hE_{ki}-\hE_{ik}\hE_{kj}\hE_{ji}
\right)
\,=\,
\f{-i}{4}\,\left(
\hF^\dagger_{ij}\hE_{kj}\hF_{ik}-\hF^\dagger_{ik}\hE_{jk}\hF_{ij}
\right)\,.
\ee
It might be interesting to study the action and spectrum of each of these combinations of $E$ and $F$ operators separately.
\footnotetext{One issue in order to extract a square-root of the squared volume operator $\hU$ in the spinorial framework is that it is of order 6 in the spinor components i.e in the creation and annihilation operators $a_i^{A}$, and that there does not exist any cubic gauge-invariant combinations of these operators. It might thus be interesting to identify another operator from which to extract the volume (squared or not) of order 8 for example, from which we could maybe extract a square-root of order 4 in the spinor components.}

One can also generalize the construction of such gauge-invariant grasping operators of higher order. The operators $E$ and $F$ are already $\SU(2)$ invariant, so we only have to deal with enforcing the  $\U(1)$ invariance. This is achieved by requiring that matching the indices i.e requiring that each edge appear the same number of times through creation operators and  annihilation operators.


\section{Spinorial Representation of the Wilson loop}

The second important class of operators in loop quantum gravity are the Wilson loop operators $\hat{W}_\cL$ which, in contrast to the grasping operators, capture non-local information about the states in $\cH_{\Gamma}$. In terms of holonomies the Wilson loop operators are defined as
\be
\hat{W}_\cL := \tr (\prod\limits_{e \subset \cL}^{\rightarrow} \hat{g}_e)
\ee
that is the trace over the oriented product of holonomies ordered along all edges $e$ part of a (oriented) loop $\cL$ (i.e. we take the inverse of a group element if the edge is oriented in the opposite direction than the loop).
An expression for classical Wilson loops in terms of the classical observables $E$ and $F$ was given in \cite{spinor}. It was however unclear how to quantize these expressions due to the ambiguity in regularizing the inverse norm factors in the holonomies at he quantum level. Having obtained an explicit form of the holonomy operator on a single edge (\ref{g_op_1}) in section \ref{formula}, we are now able to provide an explicit formula for the Wilson loop operators in terms of the generalized ladder operators $(\hat{E}, \hat{F})$.

\begin{figure}[h]
\begin{center}
\includegraphics[height=50mm]{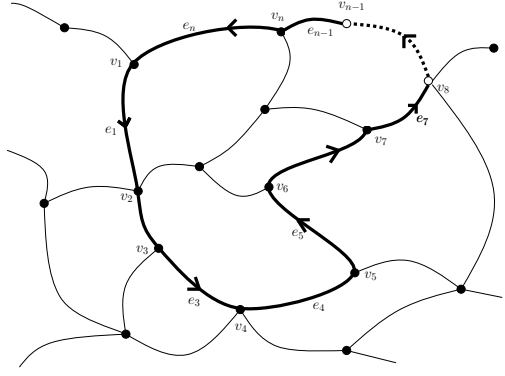}
\caption{The loop $\cL=\{e_1,e_2,..,e_n\}$ on the graph $\Gamma$.
\label{inaki}}
\end{center}
\end{figure}

Writing the group elements $g_e$ in terms of the spinor variables as in eqn.\eqref{group_element}, we use our operator ordering \eqref{g_op_1} and we regroup the creation and annihilation operations around the vertices along the loop as was done in \cite{spinor} at the classical level. This gives:
\beq
\hat{W}_\cL
&=&
\sum\limits_{r_i = 0,1} (-1)^{\sum\limits_i r_i} \cW_\cL^{\{r_i\}}  \times \prod\limits_i [2\hat{E}_{ii} +1]^{-1}\,\,, \\
\cW_\cL^{\{r_i\}}
&\equiv&
\prod\limits_i r_{i-1}r_i \hat{F}^i_{i,i-1} + (1-r_{i-1})r_i \hat{E}^i_{i-1,i} + r_{i-1}(1-r_i)\hat{E}^i_{i, i-1} + (1-r_{i-1})(1-r_i) (\hat{F}^i_{i,i-1})^\dagger\,\,.
%
\eeq
The index $i$ labels the edge around the loop (from an arbitrary origin vertex), as illustrated on figure \ref{inaki}.
Here we have chosen all the edges oriented in the same direction along the loop for the sake of simplicity. In the general case, we would get an overall sign for each edge oriented in the opposite direction.
For given $r_i$'s, we call the operator $\cW_\cL^{\{r_i\}}$ the generalized holonomy operator (following the classical nomenclature introduced in \cite{spinor}). It is a polynomial operator in the operators $\hE$ and $\hF$. Its action is fairly simple despite its seemingly complicated structure. It raises the spin by $\f12$ on the edge $i$ if $r_i=0$ and lowers it by one half if $r_i=1$.
The non-polynomial part, given classically by the inverse norm factors, is regularized by a contribution of $[2\hat{E} + 1]^{-1}$ per edge, all of them \emph{ordered to the right}. This is the simplest scenario, instead of the possibility of these inverse norm factors entering the generalized holonomies and messing up their structure. The slight difference from the ordering conjectured in \cite{spinor}, with the inverse norm factor split as a square-root on the right and one of the left, is actually due to the non-trivial factor $1/\sqrt{2j+1}$ in the map $\cT$ given in eqn.\eqref{map_wigner} between the Wigner matrices $D^j_{mn}(g)$ and their holomorphic counterpart.

\begin{figure}[h]
\begin{center}
	\psfrag{gamma}{$\Gamma$}
  \psfrag{z1}{$a, b$}
	\psfrag{z2}{$\ta, \tb$}
	\psfrag{e}{$e$}
	\psfrag{v}{$v$}
	\includegraphics[scale=.8]{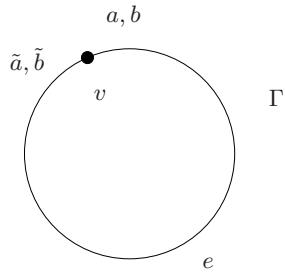}
	\caption{ \label{oneloop}  The simplest case: $\Gamma$ contains just one vertex $v$ and a loop attached to it. The associated Hilbert space $\cH_\Gamma$ consists of holomorphic square integrable functions in two spinors $\ket{z_1}$ and $\ket{z}_2$. Thus, there are four sets of ladder operators $[a_1, \bar{a}_1] = [b_1, \bar{b_1}] = [a_2, \bar{a}_2] = [b_2, \bar{b_2}] = 1$ out of which Grasping - and Wilson loop-operators, which provide a complete set of $\SU(2)$ invariants, are constructed.  }
	\end{center}
\end{figure}

This operator, expressed in terms of generalized ladder operators,  is unitarily equivalent to the standard Wilson loops of loop quantum gravity. To understand its structure and action, let us give some examples. In the simplest case, the graph is just given by one loop with a single vertex (see figure \ref{oneloop}).
There are four sets of ladder operators, one doublet for each end of the edge or equivalently one doublet on each leg around the single vertex. Let us denote these by $a, b$ and $\ta,\tb$.
The Wilson loop operator $\hat{W}$ for this loop can be decomposed into $\hat{E}$- and $\hat{F}$-operators as\footnotemark:
\be
\hat{W}
\,=\,
(\hat{F}^\dagger + \hat{F} ) \left[ 2 \hat{E} +1\right]^{-1} \,,
\ee
\footnotetext{
This is the quantization of the classical formula:
$$
\tr g =\tr \frac{\ket{z}\brar{\tz} - \ketr{z}\bra{\tz}}{\sqrt{\braaket{z}{z}\braaket{\tz}{\tz}}}
=\frac{[\tz|z\ra-\la\tz|z]}{\sqrt{\braaket{z}{z}\braaket{\tz}{\tz}}}
=\frac{F+\overline{F}}{\sqrt{\braaket{z}{z}\braaket{\tz}{\tz}}}\,.
$$
}
with
$$
\hE=a^\dagger a+b^\dagger b =\ta^\dagger \ta+\tb^\dagger \tb,
\qquad
\hF=b\ta - a\tb,
\qquad
\hF^\dagger=b^\dagger\ta^\dagger - a^\dagger\tb^\dagger
\,.
$$
The operator $\hE$ gives the spin on the single edge, i.e. the area carried by that edge, while the operators $\hF$ and $\hF^\dagger$ act at the vertex and respectively decreases and increases the spin on the edge by one half. We can compute the action of these operators on our Hilbert space. Since we have a single loop, an orthogonal basis is given by the characters $\chi_j =\sum_m \cP^j_{mm}$, and we get\footnotemark:
\be
\hE \,\chi_j\, =\, j\,\chi_j,
\qquad
\hF \,\chi_j\, =\, (2j+1)\,\chi_{j-\f12},
\qquad
\hF^\dagger \,\chi_j\, =\, (2j+1)\,\chi_{j+\f12}\,.
\ee
\footnotetext{
In order to check that $\hF^\dagger$ is indeed the adjoint operator to $\hF$, one must keep in mind that the $\cP_{mn}$ are orthonormal, so that the states $\chi_j$ are not normalized but such that $\la\chi_j|\chi_j\ra=(2j+1)$. This might seem awkward since the characters are normalized with respect to the Haar measure on the $\SU(2)$-group, $\int dg\,\chi_j(g)^2=1$. However, this is actually the reason for the $1/\sqrt(2j+1)$ factor in the $\cT$-map given in eqn.\eqref{map_wigner} between the Wigner matrices $D^j_{mn}(g)$ and their holomorphic counterpart.
}
One easily check that we indeed recover the expected action of the holonomy operator due to the factor $(2j+1)^{-1}$, as given in eqn.\eqref{hol_action}:
\be
\hat{W}\,\chi_j
\,=\,
\chi_{j-\f12}+\chi_{j+\f12}
\,=\,
\what{\chi_{\f12}(g)}\,\chi_j\,.
\ee
One can go further and check that indeed $\hat{F}^\dagger + \hat{F}=\hat{g}_{--}+\hat{g}_{++}$ with the group element operators given earlier in eqn.\eqref{g_op_1}.

\smallskip

\begin{figure}
\begin{center}
	\psfrag{gamma}{$\gamma$}
  \psfrag{1}{$v$}
	\psfrag{2}{$w$}
	\psfrag{a}{$a_1, b_1$}
	\psfrag{b}{$\ta_1, \tb_1$}
	\psfrag{c}{$\ta_2, \tb_2$}
	\psfrag{d}{$a_2, b_2$}
	\psfrag{e_1}{$e_1$}
	\psfrag{e_2}{$e_2$}
	\includegraphics[scale=.8]{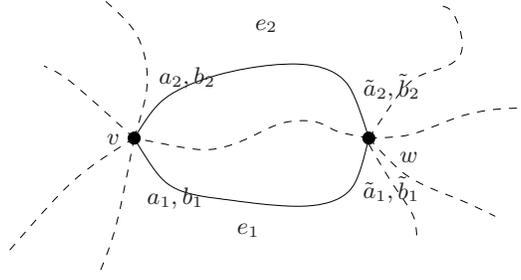}
\caption{ \label{2vertex} A loop within $\Gamma$ that goes through only two vertices. The doublet of harmonic oscillator operators $(a,b)$ is attached to the vertex $v$ while   The doublet of harmonic oscillator operators $(\ta,\tb)$ is attached to the vertex $w$.}
\end{center}
\end{figure}

Beyond this consistency check on the single loop, a more generic example is given by the following situation: consider a loop $\cL$ within a graph $\Gamma$ that goes through only 2 vertices (see figure \ref{2vertex}).
Computing the Wilson loop operator around the loop indicated in figure
we obtain\footnotemark:
\footnotetext{
This is the quantization of the classical formula:
$$
\tr g_1g_2^{-1}
=\tr \frac{\left(\ket{z_1}\brar{\tz_1} -\ketr{z_1}\bra{\tz_1}\right)\left(|\tz_2]\la z_2|-|\tz_2\ra[z_2|\right)}
{\sqrt{\braaket{z_1}{z_1}\braaket{\tz_1}{\tz_1}}\sqrt{\braaket{z_2}{z_2}\braaket{\tz_2}{\tz_2}}}
=\frac{E_{12}\tE_{12}+E_{21}\tE_{21}-F_{21}\tF_{12}-\bar{F}_{12}\overline{\tF}_{21}}
{\sqrt{\braaket{z_1}{z_1}\braaket{\tz_1}{\tz_1}}\sqrt{\braaket{z_2}{z_2}\braaket{\tz_2}{\tz_2}}}
\,.
$$
}
\be
\hat{W}
\,=\,
\left[
 \hat{E}_{12}\hat{\tE}_{12}
+ (\hat{E}_{12})^\dagger (\hat{\tE}_{12})^\dagger
+(\hat{F}_{12})^\dagger (\hat{\tF}_{12})^\dagger
+ \hat{F}_{12}\hat{\tF}_{12}
\right]\,
\left[ 2\hat{E}_{11}  + 1\right]^{-1} \left[ 2 \hat{E}_{22} + 1\right]^{-1}  \, .
\ee
This reproduces exactly the holonomy operator derived in \cite{2vertex} on the 2-vertex graph using the explicit Clebsh-Gordon coefficients.

\smallskip

In general, these generalized holonomy operators around the loop $\cL$ allow to split the full holonomy operator into smaller polynomial operators which act by simple shifts on all edges of the loop. In some simple cases, they have already been used to generate recursion relations on spin network evaluations (and more particularly on the 6j-symbol of the recoupling theory of spins) and to generate the action of the Hamiltonian constraints in 2+1 Riemannian gravity \cite{recursion_spinor}. We hope that this reformulation of all loop quantum gravity gauge-invariant operators in terms of the ladder operators $\hE$ and $\hF$ will somewhat allow a more systematic approach to the study of gauge-invariant operators entering the Hamiltonian (constraint) for loop quantum gravity in 3+1 dimensions.

\section{Outlook and Conclusion}

In this short note, we have constructed the holonomy-flux operators in the spinor representation of loop quantum gravity. Holonomies and fluxes emerge as composite operator built from a set of harmonic oscillator operators which are considered to be the more elementary operators in this picture. Because of this compositeness, there are certain operator ordering ambiguities which need to be investigated. Here we showed that an operator ordering is selected by the requirement that the classical holonomy-flux algebra be represented non-anomalously on the spinorial Hilbert space $\cH^{\rm spin}$. This guarantees that this representation of the holonomy-flux algebra is unitarily equivalent to the standard one on $L^2(\SU(2))$.

Taking $\SU(2)$-gauge invariance at the nodes of $\Gamma$ into account we constructed the familiar grasping and Wilson loop operators on $\cH^{\rm spin}$. They can be written in terms of the generalized ladder operators $\hat{E}, \hat{F}$ introduced in the $\U(N)$-formalism \cite{un1,un2,un3,spinor,2vertex}, capturing the gauge-invariant content of the individual intertwiner spaces. An interesting point to note is that in the spinor formalism the distinction between Wilson loops on the one hand and grasping operators on the other side becomes blurry: both are gauge invariant combinations of the same elementary $\hat{E}, \hat{F}$, the only difference being that the grasping operators are localized around a vertex of $\Gamma$ whereas the Wilson loop operators contain non-local information on the spin network state.

An interesting side results of this note is the observation that the spinor formalism can be used as a simple toy model to test the quantization procedure leading to the Hamiltonian constraint operator in loop quantum gravity. This quantization procedure rests on a peculiar Poisson-identity, which is used to get rid of potentially diverging operators and can be modeled by a similar (at least in spirit) Poisson-identity in the spinor formalism. Here we showed that a quantization of the holonomy operator based on that Poisson-identity leads to an anomalous representation of the holonomy flux algebra on $\cH^{\rm spin}$. While this calculation does not allow a direct conclusion for the full theory, we suggest that more toy models of this kind should be considered to collect (counter-?) evidence for an anomaly-free implementation of the Dirac algebra in loop quantum gravity.

\section*{Acknowledgments}

EL and JT acknowledge support from the Programme Blanc LQG-09 from the ANR (France).



\end{document}